\documentclass[article,pra]{revtex4}
\usepackage{color}
\usepackage{graphicx}
\usepackage{bm}
\usepackage{amsmath}
\usepackage{amsbsy}
\usepackage{hyperref}
\usepackage{enumerate}
\usepackage{upgreek}
\usepackage[subnum]{cases}
\usepackage{dsfont}

\newcommand{\tr}{ \text{tr} }

\newcommand{\pa}{ \partial }
\newcommand{\hb}{ \hbar }
\newcommand{\si}{ \sigma }

\newcommand{\ga}{ \gamma }
\newcommand{\la}{ \langle }
\newcommand{\ra}{ \rangle }
\newcommand{\del}{ \delta }
\newcommand{\al}{ \alpha }

\newcommand{\re}{ \text{Re} }
\newcommand{\im}{ \text{Im} }

\newcommand{\MB}{ \text{MB} }

\begin{document}

\title{Modular Variables and the Limits of Phase Detectability in Open Quantum Systems}

\author{S. V. Mousavi}
\email{vmousavi@qom.ac.ir}
\affiliation{Department of Physics, University of Qom, Ghadir Blvd., Qom 371614-6611, Iran}
\begin{abstract}

Modular variables serve as a striking example of quantum nonlocality, particularly in superpositions of wave packets that are spatially well separated, where the relative phase between components cannot be accessed through conventional local measurements. In this work, we explore the time evolution of Hermitian modular operators for Gaussian wave-packet superpositions under the influence of a uniform gravitational field. We consider both unitary dynamics governed by the Schrödinger equation and open-system dynamics described by the Caldeira-Leggett master equation in the high-temperature limit. Adopting the Bohmian interpretation of quantum mechanics, we compute local expectation values of these modular operators along individual particle trajectories. Our analysis shows that gravitational acceleration induces a time-varying modular signal, the expectation value of the modular observable, that remains sensitive to the relative phase between the separated wave packets. In contrast, standard local quantities such as the probability density and probability current, while modified by gravity, become insensitive to the relative phase in the regime of negligible spatial overlap. For a pair of particles coupled to a shared environment, we find that environment-induced correlations can modify the local modular expectation value observed for one particle, yielding a clear signature of environmental influence. However, the transfer of phase sensitivity via environment-generated entanglement to the modular signal of the distant particle remains negligible within the regime considered. We further demonstrate that conventional measures of coherence and entanglement do not capture the relative phase information in this non-overlapping regime. Taken together, these findings suggest that modular variables provide a useful and sensitive probe for detecting nonlocal phase information in open quantum systems.

\end{abstract}

\maketitle

{\bf{Keywords}}: Modular variables; Nonlocal observables; Bohmian mechanics; Relative phase detection; Open quantum systems; Common environment

\section{Introduction}

The relative phase between components of a quantum superposition is a fundamental physical parameter, yet its detectability depends crucially on the spatial structure of the state. In particular, for a superposition of the form $|\Psi\rangle = |\psi_1\rangle + e^{i\alpha} |\psi_2\rangle$ with non-overlapping spatial profiles ($ \psi_1(x) \psi_2(x) \approx 0 $), the phase $\alpha$ becomes inaccessible to a broad class of physically relevant quantities. Standard local or quasi-local quantities such as the probability density and the probability current density, probe only a narrow neighborhood of the diagonal of the density matrix and therefore do not access the phase-dependent cross terms connecting spatially disjoint regions. As a result, for any observer restricted to such quantities, the pure superposition is operationally indistinguishable from an incoherent statistical mixture. 
This insensitivity is structural rather than dynamical: it follows from their strictly local algebraic form and therefore persists independently of whether the evolution is unitary or dissipative.

To access phase information in such non-overlapping configurations, one must consider quantities that are intrinsically nonlocal in configuration space. A natural candidate is provided by finite translation operators of the form $e^{i\hat p \ell/\hbar}$ \cite{AhPePe-IJTP-1969, AhRo-book-2008}. A key property of this operator is its invariance under shifts of the momentum by integer multiples of $h/\ell$: replacing $p$ by $p - n h/\ell$ leaves $e^{i p \ell/\hbar}$ unchanged, since $e^{i n 2\pi}=1$. Consequently, the translation operator does not depend on the full momentum $p$, but only on its value modulo $h/\ell$. This observation motivates the definition of the modular momentum $p_{\rm mod}$, which takes values in the interval $[0,h/\ell)$. Modular variables are thus associated with periodic functions of momentum, and the operator $e^{i\hat p \ell/\hbar}$ may be regarded as the exponential of the modular momentum. Since the modular variable $e^{i\hat p_{\rm mod} \ell/\hbar}$ coincides with the translation operator, accessing modular momentum information is operationally equivalent to evaluating expectation values of $e^{i\hat p \ell/\hbar}$. For this reason, throughout the remainder of this work we work directly with the translation operator and its Hermitian counterpart, while the underlying physical content is that of modular momentum.

The nonlocality associated with modular variables is conceptually distinct from the more familiar Bell-type nonlocality. Bell nonlocality is a kinematical property of multipartite quantum states, manifested through correlations between spatially separated measurements and necessarily involving entanglement. By contrast, modular variables probe dynamical nonlocality through finite translations that encode the relative phase of spatially disjoint wave packets. This form of nonlocality does not rely on entanglement or on measurement correlations between distant subsystems, and can arise already at the single-particle level. 

Modular variables have been extensively studied in a variety of foundational and experimental contexts. In \cite{ToAhCaKaNu-NJP-2010}, a Heisenberg-picture analysis combined with modular variables and weak measurements provided a novel account of multi-slit interference in terms of nonlocal interactions between a localized particle and distant slits. 
In Ref.~\cite{To-JPCS-2007}, several connections are established between superoscillations, weak values, and modular variables, demonstrating how uncertain phases in superoscillatory projections generate high-momentum components, and how definite phases lead to localization via nonlocal exchange of modular variables, with nonlocality further quantified via weak measurements.
Reference \cite{LoAhToBeAs-QS-2014} further clarified the conceptual foundations of modular variables, arguing that they are naturally embedded in a quantum phase-space framework closely related to Schwinger's discrete kinematics, and highlighting the distinction between kinematical and dynamical nonlocality. On the experimental side, modular-variable entanglement has been demonstrated in multi-slit photon experiments \cite{CaFeBoAsPaWa-PRA-2012}, where entropy-based witnesses were shown to outperform variance-based criteria in detecting entanglement and steering. Modular variables have been used to characterize interference and entanglement in multiparticle wave packets \cite{BaFaKeCoMiWa-PRA-2015}, revealing a complementarity between particle number and momentum peak width that enables the detection of genuine multipartite entanglement. These works establish modular variables as physically meaningful quantities beyond idealized interference scenarios. In Ref.~\cite{KeKeWaCoMi-PRA-2016}, modular variables enable fault-tolerant binary encoding in infinite-dimensional Hilbert spaces while relaxing the unphysical requirement of infinite phase-space localization for logical states, allowing encoded discrete information to be read out via modular measurements, an approach experimentally implemented using the transverse degrees of freedom of single photons.

In realistic settings, quantum systems are subject to environmental interactions, raising the question of how modular phase information behaves under open-system dynamics. While decoherence typically suppresses interference effects in local quantities, it is not evident whether environmental noise destroys the phase information encoded in modular variables or merely reduces its observability. Addressing this issue requires a dynamical analysis that treats unitary and dissipative evolution on equal footing. In this work, we investigate the behavior of modular variables under both Schr\"odinger evolution and dissipative dynamics described by the Caldeira-Leggett (CL) equation in the high-temperature regime \cite{CaLe-PA-1983, Ca-book-2014}. For a recent work on different aspects of this formalism see \cite{MoMi-EPJP-2022}. 
Beyond the modular-variable framework, the degradation of phase-encoded quantum information due to phase diffusion and noise has also been extensively studied in quantum phase communication channels \cite{TrTeOlPa-PRA-2015, TeTrOlPa-PS-2015}. Here, however, we adopt a different perspective by focusing on modular variables within the CL framework.

To gain additional physical insight, we employ Bohmian mechanics (BM) as an interpretational framework \cite{Bo-PR-1952, DuGoZa-JSP-1992, Holland-book-1993}. BM offers a clear separation between particle trajectories and wave-function structure, allowing one to analyze the role of empty waves without modifying quantum predictions. 
Recent tunneling studies have suggested apparent contradictions with BM \cite{ShPuMaToKl-Nat-2025}. However, \cite{Ni-arXiv-2025} and \cite{WaWaWaLu-arXiv-2025} argue that these discrepancies stem from improperly defined velocities. Additionally, \cite{DrLaNa-arXiv-2025} further supports the internal consistency of BM by clarifying the limitations of stationary treatments in the evanescent regime and highlighting the role of radiative leakage in shaping the Bohmian velocity.  
Using Holland's notion of local expectation values \cite{Holland-book-1993}, we examine modular variables evaluated along Bohmian trajectories. While such local expectation values are not directly measurable, they serve as a valuable conceptual tool for investigating how nonlocal phase information is accessed conditionally on the particle's trajectory, and how empty-wave components though dynamically inactive still contribute to modular correlations.
We stress that modular expectation values correspond to standard quantum observables, whereas the Bohmian local expectation values used here are trajectory-dependent and serve only as interpretational tools.

We further extend our analysis to two-particle systems, including both distinguishable and indistinguishable particles, within the framework of the double CL equation with a common environment \cite{CaMoPo-PA-2010}. This extension allows us to investigate whether phase detection via modular variables necessitates genuinely multipartite observables, or whether a local modular quantity acting on a single particle remains sufficient even in the presence of environment-induced correlations. 
While previous works have explored related aspects of the CL framework such as arrival-time distributions, diffraction in time, and identical particles in dissipative settings \cite{MoMi-EPJP-2022}, as well as the dynamics of quantum correlations, coherence, entropy, and entanglement for initially squeezed states in both common and separate environments \cite{Mo-EPJP-2025, Mo-PS-2025}, and interference and diffraction phenomena within both CL and effective Hamiltonian approaches \cite{MoMi-EPJP-2020}, their focus has remained on standard observables and information-theoretic quantifiers. 
Additionally, momentum-space decoherence in the CL framework has been examined and subsequently generalized to systems of identical particles \cite{MoMi-Entropy-2021}.
In contrast, the present work is centered on modular variables and the limits of phase detectability within the CL formalism, addressing a qualitatively different question: whether relative phase information in non-overlapping superpositions can survive environmental interactions and remain operationally accessible.
Finally, we examine the limitations of standard information-theoretic quantifiers such as coherence measures, entropy-based quantities, and entanglement indicators, in detecting relative phase information in non-overlapping superpositions. We show that their insensitivity follows directly from their construction in terms of local or quasi-local quantities and is therefore structural rather than model-dependent.
%
To the best of our knowledge, however, modular variables and their local expectation values have not previously been studied within the dissipative CL framework for non-overlapping wave-packet superpositions. In this sense, gravity and dissipation primarily provide the dynamical setting of the present analysis rather than its main conceptual novelty.

The paper is organized as follows. In Section~\ref{sec: EoM-exp}, we derive the Heisenberg equation of motion for the displacement operator and its expectation value for a general wave function evolving under a uniform gravitational field, comparing unitary Schr\"odinger dynamics with the dissipative CL framework. Section~\ref{sec: nonover-GW} analyzes a superposition of two non-overlapping Gaussian wave packets under gravitational influence in both dynamical settings, examining the evolution of modular variable expectation values and developing a trajectory-resolved description within Bohmian mechanics via local expectation values. Section~\ref{sec:two_particle} extends the analysis to two-particle systems, exploring the impact of quantum statistics and common-environment interactions, and clarifying the inability of standard coherence and entanglement quantifiers to detect the relative phase. Section~\ref{sec: num-res} presents supporting numerical simulations. We summarize and conclude in Section~\ref{sec: sum-con}.

\section{Equation of motion for the expectation value of modular variable} \label{sec: EoM-exp}

In this section we derive the time evolution of the translation operator within two different dynamical frameworks. We first consider unitary evolution governed by the Schr\"odinger equation, and then extend the analysis to the CL master equation in order to incorporate the effects of dissipation and environmental decoherence.

\subsection{Schr\"odinger approach}

The time evolution of the expectation value of a time-independent operator $\hat{A}$ is governed by
\begin{eqnarray}
\frac{d}{dt} \la \hat{A} \ra_t &=& \frac{1}{i\hb} \la [\hat{A}, \hat{H}] \ra_t .
\end{eqnarray}
Applying this equation to the translation operator $e^{i \hat{p} \ell / \hb}$, one finds
\begin{eqnarray} \label{eq: heis-eq}
\frac{d}{dt} \la e^{i \hat{p} \ell / \hb} \ra_t
&=& \frac{1}{i\hb}  \la [e^{i \hat{p} \ell / \hb}, V(\hat{x})] \ra_t = \frac{1}{i\hb} 
 \la ( V(\hat{x}+\ell) - V(\hat{x}) ) e^{i \hat{p} \ell / \hb} \ra_t .
\end{eqnarray}
As expected, in the absence of any potential energy this expectation value remains constant in time.
For a uniform gravitational potential $V(\hat{x}) = m g \hat{x}$, this reduces to
\begin{eqnarray}
\frac{d}{dt} \la e^{i \hat{p} \ell / \hb} \ra_t
&=& -\frac{i}{\hb} m g \ell \la  e^{i \hat{p} \ell / \hb} \ra_t ,
\end{eqnarray}
which yields the solution
\begin{eqnarray}
\la e^{i \hat{p} \ell / \hb} \ra_t
&=& e^{-i m g \ell t / \hb} ~ \la e^{i \hat{p} \ell / \hb} \ra_0 .
\end{eqnarray}
Consequently, the expectation value of the corresponding Hermitian operator evolves as
\begin{eqnarray}
\la \cos(\hat{p} \ell / \hb) \ra_t
&=& 
\frac{ e^{-i m g \ell t / \hb} ~ \la e^{i \hat{p} \ell / \hb} \ra_0
+ e^{i m g \ell t / \hb} ~ \la e^{-i \hat{p} \ell / \hb} \ra_0 }{2} .
\end{eqnarray}

It is worth emphasizing that the operator equation of motion
\[
\frac{d}{dt} e^{i \hat{p} \ell / \hb}
= \frac{i}{\hb} \big[ V(\hat{x}) - V(\hat{x}+\ell) \big] e^{i \hat{p} \ell / \hb}
\]
is intrinsically nonlocal. Even when the local force $dV/dx$ vanishes at the particle's position, the modular variable can still evolve if $V(\hat{x}) \neq V(\hat{x}+\ell)$ \cite{AhRo-book-2008}. This feature highlights the sensitivity of modular variables to spatially separated regions of the potential.

For completeness, the same result can be obtained within the von Neumann framework where the quantum systems are described by density matrices. One has
\begin{eqnarray}
\frac{d}{dt} \la \hat{A}  \ra_t
&=& \frac{d}{dt} \tr( \hat{\rho}(t) \hat{A} ) = \tr \left( \frac{\pa \hat{\rho}(t)}{\pa t} \hat{A} \right) 
= \tr \left( \frac{ [\hat{H}, \hat{\rho}(t)] }{i\hb} \hat{A} \right) = \frac{1}{i\hb} \tr( \hat{\rho}(t) [\hat{A}, \hat{H}] )
\nonumber \\
&=& \la [\hat{A}, \hat{H}] \ra_t ,
\end{eqnarray}
where $\hat{\rho}(t)$ denotes the density operator, and we have used the von Neumann equation along with the cyclic property of the trace.

\subsection{Caldeira-Leggett approach}

We now turn to the dynamics generated by the Caldeira-Leggett master equation. The CL model provides a microscopic description of quantum Brownian motion, where dissipation arises from the interaction between a system and its environment. Typically, the environment is modeled either as a large collection of independent harmonic oscillators \cite{Ca-book-2014} or, equivalently, as a free quantum field \cite{BhHaHaPa-PRD-2023, Zu-PT-1991}.

The derivation of the CL equation involves several approximations. One assumes an initially uncorrelated state, $\rho_{SE}(0) = \rho_S(0) \otimes \rho_E(0)$, so that correlations arise solely from the interaction. Here, $\rho_S(0)$ denotes the initial density operator of the system typically taken to be a simple harmonic oscillator with frequency $\Omega$ while $\rho_E(0)$ represents the density operator of the environment. The Born-Markov approximation is then invoked: the Born approximation assumes weak system-environment coupling, while the Markov approximation assumes a short environmental correlation time, leading to a time-local (memoryless) evolution. One further adopts an Ohmic spectral density $J(\omega) \propto \omega$ (at low frequencies) with a Lorentz-Drude cutoff $\Lambda$. Within this framework, the Born--Markov approximation is valid provided that $ \hbar \gamma \ll \min\{ \hbar \Lambda, 2\pi k_B T \} $ \cite{LaMaLe-book-2019}. Finally, taking the combined limits of high temperature and large cutoff, $ k_B T / \hbar \gg \Lambda \gg \Omega $, yields the CL equation, 
\begin{eqnarray} \label{eq: CL-abs}
\frac{\pa \hat{\rho}(t)}{\pa t}
&=& \frac{1}{i\hb} [\hat{H}, \hat{\rho}(t)]
+ \frac{\ga}{i\hb} [\hat{x}, \{\hat{p}, \hat{\rho}(t)\}]
- \frac{D}{\hb^2} [\hat{x}, [\hat{x}, \hat{\rho}(t)]] ,
\end{eqnarray}
where the last term represents decoherence, appearing as a spatial diffusion term that suppresses off-diagonal coherence in the position basis \cite{LaMaLe-book-2019, Sch-PR-2019}. Here, $\ga$ is the relaxation rate (dissipation constant) and $D = 2 m \ga k_B T$ is the diffusion coefficient.
As this derivation indicates, when the system under consideration is a simple harmonic oscillator with frequency $\Omega$, the high-temperature regime is typically characterized by $ k_B T \gg \hbar \Omega $.
For a free damped particle (and similarly for a particle in a linear potential), there is no intrinsic system frequency. In this case, the relevant dynamical scale is set by the relaxation rate $\gamma$, and consistency of the Markovian high-temperature regime requires $ k_B T \gg \hbar \gamma $ \cite{Ca-book-2014, Ve-PRA-1994}.

The time derivative of the expectation value of an operator $\hat{A}$ then becomes
\begin{eqnarray}
\frac{d}{dt} \la \hat{A} \ra_t
&=& \tr \left( \frac{\pa \hat{\rho}(t)}{\pa t} \hat{A} \right) \nonumber \\
&=& \frac{1}{i\hb} \tr \left( [\hat{H}, \hat{\rho}(t)] \hat{A} \right)
+ \frac{\ga}{i\hb} \tr \left( [\hat{x}, \{\hat{p}, \hat{\rho}(t)\}] \hat{A} \right)
- \frac{D}{\hb^2} \tr \left( [\hat{x}, [\hat{x}, \hat{\rho}(t)]] \hat{A} \right) .
\end{eqnarray}
The Hamiltonian contribution coincides with the Schr\"odinger result derived above. We now evaluate the remaining two terms.

First, using $ [\hat{x}, [\hat{x}, \hat{\rho}(t)]] = \hat{x}^2 \hat{\rho}(t) - 2 \hat{x} \hat{\rho}(t) \hat{x} + \hat{\rho}(t) \hat{x}^2 $, one finds
\begin{eqnarray} \label{eq: third}
\tr \left( [\hat{x}, [\hat{x}, \hat{\rho}(t)]] \hat{A} \right)
&=& \tr \left( \hat{\rho}(t) [\hat{x}, [\hat{x}, \hat{A}]] \right) .
\end{eqnarray}
Next, for the dissipative term one has
\begin{eqnarray} \label{eq: second0}
\tr \left( [\hat{x}, \{\hat{p}, \hat{\rho}(t)\}] \hat{A} \right)
&=& \tr \left( \hat{\rho}(t) ( \hat{A} \hat{x} \hat{p}
+ \hat{p} \hat{A} \hat{x}
- \hat{x} \hat{A} \hat{p}
- \hat{p} \hat{x} \hat{A} ) \right) .
\end{eqnarray}
Using the identity
$
\hat{A} \hat{x} \hat{p}
+ \hat{p} \hat{A} \hat{x}
- \hat{x} \hat{A} \hat{p}
- \hat{p} \hat{x} \hat{A}
= 2 i \hb \hat{A} - [\hat{x}, \{\hat{p}, \hat{A}\}] ,
$
this term becomes
\begin{eqnarray} \label{eq: second}
\tr \left( [\hat{x}, \{\hat{p}, \hat{\rho}(t)\}] \hat{A} \right)
&=& \tr \left( \hat{\rho}(t) ( 2 i \hb \hat{A} - [\hat{x}, \{\hat{p}, \hat{A}\}] ) \right) .
\end{eqnarray}
Collecting all contributions, we obtain the evolution equation for the expectation value of the operator $\hat{A}$ within the CL framework as
\begin{eqnarray} \label{eq: He_CL}
\frac{d}{dt} \la \hat{A} \ra_t
&=& \frac{1}{i\hb} \la [\hat{A}, \hat{H}] \ra_t
- \frac{\ga}{i\hb} \la [\hat{x}, \{\hat{p}, \hat{A}\}] \ra_t
+ 2 \ga \la \hat{A} \ra_t
- \frac{D}{\hb^2} \la [\hat{x}, [\hat{x}, \hat{A}]] \ra_t .
\end{eqnarray}

We now specialize to operators of the form $\hat{A} = f(\hat{p})$. In this case,
\begin{eqnarray}
[\hat{x}, \{\hat{p}, f(\hat{p})\}]
&=& 2 [\hat{x}, \hat{p} f(\hat{p})]
= 2 i \hb \frac{d (p f(p))}{d p}
= 2 i \hb ( f(\hat{p}) + \hat{p} f'(\hat{p}) ) ,
\end{eqnarray}
and
\begin{eqnarray}
[\hat{x}, [\hat{x}, f(\hat{p})]]
&=& -\hb^2 f''(\hat{p}) .
\end{eqnarray}
For the displacement operator $ f(\hat{p}) = e^{i \hat{p} \ell / \hb} $, using
$ e^{i \hat{p} \ell / \hb} g(\hat{x}) = g(\hat{x}+\ell) e^{i \hat{p} \ell / \hb} $, one obtains
\begin{eqnarray}
[\hat{x}, \{\hat{p}, e^{i \hat{p} \ell / \hb}\}]
&=& 2 i \hb \left( \hat{\mathds{1}} + i \frac{\hat{p} \ell}{\hb} \right) e^{i \hat{p} \ell / \hb} ,
\end{eqnarray}
and
\begin{eqnarray}
[\hat{x}, [\hat{x}, e^{i \hat{p} \ell / \hb}]]
&=& \ell^2 e^{i \hat{p} \ell / \hb} .
\end{eqnarray}
Finally, in the presence of a uniform gravitational potential, the equation of motion for the modular variable expectation value becomes
\begin{eqnarray}
\frac{d}{dt} \la e^{i \hat{p} \ell / \hb} \ra_t
&=& 
\left( -\frac{i}{\hb} m g \ell - \frac{D \ell^2}{\hb^2} \right)
\la  e^{i \hat{p} \ell / \hb} \ra_t
- 2 \ga i \frac{\ell}{\hb}
\la \hat{p} e^{i \hat{p} \ell / \hb} \ra_t .
\end{eqnarray}
The three terms on the right-hand side have distinct physical origins:  
(i) the purely imaginary term \(-\frac{i}{\hbar} m g \ell\) arises from the commutator with the gravitational potential \(mg\hat{x}\) and introduces a constant frequency shift in the unitary evolution of the modular variable proportional to \(mg\ell/\hbar\);   
(ii) the real term \(-\frac{D \ell^2}{\hbar^2}\) describes exponential damping due to thermal diffusion (\(D = 2 m \gamma k_B T\)), which suppresses the amplitude of the modular signal;  
(iii) the last term, proportional to \(\la \hat{p} e^{i \hat{p} \ell / \hb} \ra_t\), arises from the momentum-damping term (\(-\gamma [\hat{x}, \{ \hat{p}, \hat{\rho}(t) \}\)) in the CL master equation, introducing a dissipative coupling between the modular variable and the momentum operator.  
Consequently, the environment not only damps the modular amplitude (decoherence) but also alters its oscillation frequency and phase evolution, illustrating how modular variables while sensitive to nonlocal phase information, remain subject to standard open system effects.

\section{Two non-overlapping Gaussian wave packets in a gravitational field} \label{sec: nonover-GW}

In this section, we investigate the evolution of a superposition of two localized, non-overlapping Gaussian wave packets of equal width, initially centered symmetrically about the origin with a center-to-center separation $L$, in the presence of a uniform gravitational field. The left wave packet is prepared with zero initial momentum, while the right wave packet is assigned a non-zero initial kick momentum. The dynamics are analyzed within two distinct frameworks: the unitary Schr\"odinger evolution and the dissipative CL dynamics in the high-temperature limit, which are discussed in separate subsections. Within this setting, we introduce and evaluate the local expectation value of the modular variable in the Bohmian mechanical framework.

\subsection{The Schr\"odinger framework} 

We begin by reviewing the dynamics of a single Gaussian wave packet in a uniform gravitational field. At the initial time $t=0$, the wave function is chosen as
\begin{eqnarray} \label{eq: gauss0}
\phi(x, 0) &=& \frac{1}{(2\pi \si_0^2)^{1/4}}
\exp \left[ - \frac{(x-x_0)^2}{4 \si_0^2} + i \frac{p_0}{\hb} (x-x_0) \right],
\end{eqnarray}
where $\si_0$ denotes the initial width, $x_0$ the initial center, and $p_0$ the initial (kick) momentum. The time evolution is governed by the Schr\"odinger equation
\begin{eqnarray} \label{eq: Sch-eq}
i \hb \frac{\pa}{\pa t} \phi(x, t)
&=& \left( - \frac{\hb^2}{2m} \frac{\pa^2}{\pa x^2} + m g x \right) \phi(x, t),
\end{eqnarray}
whose exact solution can be written as
\begin{eqnarray} \label{eq: gauss-grav}
\phi(x, t)
&=& \frac{1}{ (2\pi s_t^2)^{1/4} }
\exp \left[
- \frac{ (x-x_t)^2 }{4 s_t \si_0}
+ \frac{i}{\hb} p_t (x-x_t)
+ \frac{i}{\hb} \mathcal{A}_t
\right].
\end{eqnarray}
Here $s_t$ is the complex width, $x_t$ denotes the classical trajectory of the wave packet center, and $\mathcal{A}_t$ is the classical action, given explicitly by
\begin{numcases}~
s_t = \si_0 \left( 1 + i \frac{\hb t}{ 2m \si_0^2 } \right), \\
x_t = x_0 + \frac{p_0}{m} t - \frac{1}{2} g t^2, \label{eq: xt-Sch}\\
\mathcal{A}_t
= \int_0^t dt' \left( \frac{1}{2} m \dot{x}_t^2 - m g x_t \right)
= \left( \frac{p_0^2}{2m} - m g x_0 \right) t
- p_0 g t^2 + \frac{1}{3} m g^2 t^3 ,
\end{numcases}
with $ p_t = m \dot{x}_t = p_0 - m g t $.

From the wave function \eqref{eq: gauss-grav}, one directly obtains the probability density and probability current density (PCD),
\begin{eqnarray}
\varrho(x, t)
&=& |\phi(x, t)|^2
= \frac{1}{\sqrt{2\pi}~ \si_t}
\exp \left[ - \frac{ (x-x_t)^2 }{2 \si_t^2 } \right],
\label{eq: rho-gauss} \\
j(x, t) &=& \frac{\hb}{m} \im \left\{ \phi^*(x, t) \frac{\pa}{\pa x} \phi(x, t) \right\}
\equiv \rho(x, t) v(x, t),
\label{eq: cur-gauss}
\end{eqnarray}
where the time-dependent width is
\begin{eqnarray} \label{eq: sigmat}
\si_t
&=& \si_0 \sqrt{ 1 + \frac{\hb^2 t^2}{4m^2 \si_0^4} } .
\end{eqnarray}
and we have introduced the Bohmian velocity field via Eq.~\eqref{eq: cur-gauss} as the ratio of PCD to the probability density. For the Gaussian wave packet given in Eq.~\eqref{eq: gauss-grav}, the corresponding Bohmian velocity field is then given by
\begin{eqnarray} \label{eq: bm-vel-field}
v(x, t)
&=& \frac{ 8 m \si_0^4 p_0
+ ( 2 \hb^2(x-x_0) - 8 m^2 g \si_0^4 ) t
- g \hb^2 t^3 }
{ 8 m^2 \si_0^2 \si_t^2 } .
\end{eqnarray}
Integration of this velocity field yields the Bohmian trajectory
\begin{eqnarray} \label{eq: BM-traj-Sch}
X(t)
&=& x_t + (X_0 - x_0) \frac{\si_t}{\si_0},
\end{eqnarray}
where $X_0$ is the initial Bohmian position.

Here it is worth noting that the above introduction of BM as a first-order theory is commonly referred to as the minimal formulation \cite{DuGoZa-JSP-1992}. Alternatively, BM can also be formulated as a second-order theory \cite{Bo-PR-1952, Holland-book-1993}. In this approach, one inserts the polar decomposition of the wave function into the Schr\"odinger equation, which then separates into two coupled equations: the real part yields a modified Hamilton-Jacobi equation containing an additional term known as the quantum potential, while the imaginary part gives rise to the usual continuity equation. Defining the Bohmian velocity field as the gradient of the phase of the wave function, these two equations together lead to a second-order, Newtonian-like equation of motion that incorporates quantum corrections.

\medskip

We now turn to a superposition of two initially non-overlapping Gaussian wave packets,
\begin{eqnarray} \label{eq: wave0-sup}
\Psi(x, 0)
&=& N \frac{ \psi_A(x, 0) + e^{i \alpha} \psi_B(x, 0) }{\sqrt{2}},
\end{eqnarray}
with
\begin{eqnarray}
\psi_A(x, 0)
&=& \phi(x, 0) \bigg |_{x_0=-L/2,\, p_0=0}, \\
\psi_B(x, 0)
&=& \phi(x, 0) \bigg |_{x_0=L/2,\, p_0=\hb k}.
\end{eqnarray}
Both packets have equal initial width $\si_0$ and are centered at $x=\pm L/2$, while only the right packet receives a momentum kick $\hb k$.
The normalization constant is given by
\begin{eqnarray} \label{eq: normalization}
N &=& \frac{1}{\sqrt{ 1 + \re\{ e^{i\alpha} \la \psi_A | \psi_B \ra \} }}
= \left( 1 + \cos( \alpha - k L /2 ) \exp \left[ - \frac{L^2}{8\si_0^2} - \frac{1}{2} k^2 \si_0^2 \right] \right)^{-1/2}.
\end{eqnarray}
In the limit $L \gg \si_0$, the overlap between the two packets becomes negligible and $N \approx 1$.
For a superposition of two spatially separated wave packets with negligible spatial overlap, all local observables are insensitive to the relative phase $\alpha$. In contrast, the expectation value of the displacement operator $e^{i \hat{p}L/\hbar}$, as well as its Hermitian counterpart $\cos(\hat{p}L/\hbar)$, depends explicitly on $\alpha$, thereby revealing the intrinsically nonlocal nature of these operators. These operators are invariant under the replacement $\hat{p} \to \hat{p} - n\,2\pi\hbar/L$, where $n$ is an integer. For this reason, the associated quantity is referred to as the modular momentum \cite{AhRo-book-2008}.

Under the condition of negligible overlap between the wave packets, the expectation value of the Hermitian modular variable takes the form
\begin{eqnarray} \label{eq: expval-mod}
\la \cos(\hat{p}L/\hb) \ra
&=& \frac{1}{2} e^{-k^2 \si_0^2/2}
\cos \left[ \alpha - \frac{m g L}{\hb} t \right],
\end{eqnarray}
exhibiting oscillations with period $ \uptau = \dfrac{2\pi \hb}{m g L} $. 

In Bohmian mechanics, the ontology consists of both the particle position and the wave function, which evolves according to the Schrödinger equation. The particle trajectory is guided by the phase of the wave function through the Bohmian velocity field, while the quantum potential encodes the influence of the wave amplitude.
When the wave function splits into non-overlapping components, the particle occupies only one branch. The remaining branch is referred to as an \emph{empty wave} \cite{Holland-book-1993}. Empty waves evolve independently and remain part of the total wave function, but do not influence the particle's motion when their support is disjoint from the particle position.
It is crucial to emphasize that empty waves exert no force, do not modify the Bohmian velocity field, and do not affect the quantum potential in the non-overlapping regime.

Following Holland \cite{Holland-book-1993}, we define the local expectation value of an operator $\hat B$ as
\begin{equation}
B_{\rm loc}(x,t) = \re \left\{ \frac{ \Psi^*(x,t) \hat B \Psi(x,t) }{|\Psi(x,t)|^2} \right\}.
\end{equation}
This quantity is not an observable and does not enter the Bohmian equations of motion. Its role is purely interpretative: it provides a spacetime-resolved decomposition of the global expectation value,
\begin{equation}
\langle \hat B \rangle = \int dx\, |\Psi(x,t)|^2 B_{\text{loc}}(x,t).
\end{equation}
The local expectation value coincides with the real part of the weak value of $\hat B$ post-selected on position. This connection shows that the information extracted by $B_{\text{loc}}$ is not an artificial Bohmian construct, but closely related to quantities with operational meaning in standard quantum mechanics.

Along a Bohmian trajectory given by \eqref{eq: BM-traj-Sch}, the local expectation value of the Hermitian modular variable $ \cos(\hat{p}L/\hb) $ reduces to
\begin{eqnarray} \label{eq: expval-traj}
A(X(t), t)
&=& \frac{1}{2} \exp \left[ \frac{ \hb k t }{ 2 m \si_t^2 } \left( - \frac{ \hb k t }{2 m} + \left( X_0 + \frac{L}{2} \right) \frac{\si_t}{\si_0} \right) \right] \nonumber \\ 
&\times&
\cos \left[
\alpha
+ \frac{1}{2}
\left(
- \frac{2 m g L t}{\hb}
- \frac{ \hb k^2 \si_0^2}{ m \si_t^2 } t
+ \frac{k(L+2X_0)\si_0}{\si_t}
\right)
\right].
\end{eqnarray}

\subsection{The Caldeira-Leggett framework}

We now consider the evolution of the same system within the CL framework. In the position representation and in the presence of a uniform gravitational potential $V(x)=mgx$, the master equation \eqref{eq: CL-abs} becomes
\begin{eqnarray} \label{eq: CL-1p-gravity}
\frac{\pa \rho}{\pa t}
&=& \left[
- \frac{\hb}{2mi}\left( \frac{\pa^2}{\pa x^2} - \frac{\pa^2}{\pa y^2} \right)
- \ga (x-y) \left( \frac{\pa}{\pa x} - \frac{\pa}{\pa y} \right)
- \frac{D}{\hb^2} (x-y)^2
+ \frac{ mg (x-y) }{i \hb}
\right] \rho(x,y,t).
\end{eqnarray}
Introducing new coordinates $(r, R) = (x-x', (x+x')/2)$, one obtains
\begin{eqnarray} \label{eq: CL1p-rR}
\frac{\pa \rho}{\pa t}
&=& \left[
\frac{ i \hb}{m} \frac{\pa^2}{\pa r \pa R}
- 2 \ga r\frac{\pa}{\pa r}
- \frac{D}{\hb^2} r^2
+ \frac{ m g }{i \hb} r
\right] \rho(r, R, t).
\end{eqnarray}
This equation admits a continuity-equation interpretation with probability current
\begin{eqnarray} \label{eq: cur-CL}
j(x, t)
&=& \frac{\hb}{m} \im\left\{
\frac{\pa \rho(x, y, t)}{\pa x}\bigg|_{y=x}
\right\}.
\end{eqnarray}
The evolution of the initial Gaussian state \eqref{eq: gauss0} preserves the functional form of the probability density and current, as in the unitary Schr\"odinger dynamics, while dissipation and diffusion modify the center and width according to
\begin{numcases}~
x_t = x_0 + \frac{p_0}{m} \tau(t) - \frac{ g }{2 \ga } (t - \tau(t)),  \label{eq: xt-CL} \\
w_t = \sqrt{
\si_0^2 \left( 1 + \frac{\hb^2 \tau^2(t)}{4m^2 \si_0^4} \right)
- \frac{ 3 + e^{-4\ga t} - 4 e^{-2\ga t} - 4 \ga t } 
{ 8 m^2 \ga^3 } D }, \label{eq: width-CL} 
\end{numcases}
with
\begin{eqnarray} \label{eq: tau-t}
\tau(t) &=& \frac{1-e^{-2\ga t}}{2\ga}.
\end{eqnarray}
The corresponding Bohmian trajectories are given by
\begin{eqnarray} \label{eq: BM-traj-CL}
X(t) &=& x_t + (X_0 - x_0) \frac{w_t}{\si_0}.
\end{eqnarray}

For the expectation value of the momentum operator and any well-defined function thereof, one finds
\begin{eqnarray} 
\la \hat{p} \ra_t
&=& \int dR ~ \frac{\hb}{i} \frac{\pa \rho}{\pa r}\bigg|_{r=0},
\label{eq: mom-exp}
\end{eqnarray}
and
\begin{eqnarray} 
\la f(\hat{p}) \ra_t
&=& \int dr \int dR ~ \del(r) ~
f\left( \frac{\hb}{i} \frac{\pa}{\pa r} \right)
\rho(R+r/2, R-r/2, t).
\end{eqnarray}
In particular, the expectation value of the displacement operator becomes
\begin{eqnarray} \label{eq: tr-op-exp} 
\la e^{i \hat{p} L / \hb} \ra_t
&=& \int dx' ~ \rho(x'+L, x', t),
\end{eqnarray}
leading to
\begin{eqnarray} \label{eq: mod-op-exp} 
\la \cos(\hat{p}L/\hb) \ra_t
&=& \int dx' ~ \frac{ \rho(x'+L, x', t) + \rho(x'-L, x', t) } {2}.
\end{eqnarray}
From \eqref{eq: tr-op-exp} we define the local expectation value of the displacement operator as
\begin{eqnarray} \label{eq: tr-op-loc-exp}
\mathcal{T}_L(x, t)
&=& \frac{ \rho(x+L, x, t) }{ \rho(x, x, t) } .
\end{eqnarray}

Solution of the master equation \eqref{eq: CL1p-rR} via the method of characteristics for the initial superposition state \eqref{eq: wave0-sup},
\begin{eqnarray} \label{eq: rho0}
\rho_0(x, x')
&=& \frac{ \psi_A(x, 0) + e^{i \alpha} \psi_B(x, 0) }{\sqrt{2}}
\frac{ \psi^*_A(x', 0) + e^{-i \alpha} \psi^*_B(x', 0) }{\sqrt{2}},
\end{eqnarray}
has been presented in appendix \ref{sec: appendix}. Using this solution one finally obtains
\begin{eqnarray} \label{eq: expval-mod-1p}
\la \cos(\hat{p}L/\hb) \ra_t
&=& \frac{1}{2}
\exp\left[
- \frac{D L^2}{2\hb^2 \ga} e^{-2\ga t} \sinh(2\ga t)
- \frac{L^2}{2\si_0^2} e^{-2\ga t} \sinh^2(\ga t)
- \frac{k^2 \si_0^2}{2}
\right] \nonumber \\
&\times&
\cos \left[
\alpha - L \frac{ 1-e^{-2\ga t} }{2}
\left( k + \frac{mg}{\hb \ga} \right)
\right],
\end{eqnarray}
provided that the two wave packets remain non-overlapping.
This result makes explicit that thermal fluctuations lead to a damping of the modular oscillation amplitude, whereas the phase is insensitive to thermal noise and is modified only through dissipative effects.

\section{Modular Phase Detection in Two-Particle Systems} \label{sec:two_particle}

We extend our analysis to two-particle systems to examine whether quantum statistics and environment-induced correlations influence the detectability of the relative phase in superpositions of spatially non-overlapping wave packets. We demonstrate that the central result remains robust: a modular observable acting locally on a single particle provides a sufficient witness of the relative phase, even for indistinguishable bosons or fermions. This confirms that phase detectability originates from the intrinsic nonlocal structure of the modular operator rather than from global multi-particle observables or multipartite correlations. In contrast, standard quantum coherence and entanglement quantifiers, which rely on reduced density matrices, remain insensitive to the relative phase in this configuration. This contrast highlights that the relevant phase information is encoded in the nonlocal algebraic properties of modular variables rather than in conventional measures of quantum correlations.

Consider a bipartite system with the following single-particle states: a spatial superposition of two non-overlapping states $ |\psi_A \ra $ and $ |\psi_B \ra $ with a relative phase $\alpha$, and a well-localized state $ | \chi \ra $. The distance between the corresponding non-overlapping well-localized wave packets $\psi_A(x)$ and $\psi_B(x)$ is taken to be $L$, with $\psi_B(x)$ located to the right of $\psi_A(x)$. The two-particle state is taken initially separable.

When the particles are distinguishable, obeying classical Maxwell--Boltzmann (MB) statistics, the total initial state is given by:
\begin{equation}
| \Psi(0) \ra_{\MB} = \frac{ | \psi_A \ra_1 + e^{i\alpha} | \psi_B \ra_1 }{ \sqrt{2} } \otimes | \chi \ra_2 .
\end{equation}
The single-particle modular variable $e^{i \hat{p}_1 L / \hbar}$ acts only on particle $1$. Its expectation value becomes
\begin{eqnarray}
\langle e^{ i \hat{p}_1 L / \hb } \rangle &=& \tr_1 \big(  e^{ i \hat{p}_1 L / \hb } \tr_2( | \Psi(0) \ra_{\MB}\la \Psi(0) | ) \big)
= \frac{1}{2}  e^{i\alpha} \langle \psi_A | e^{ i \hat{p}_1 L / \hb } | \psi_B \rangle .
\end{eqnarray}
Crucially, only the cross term $\langle \psi_A | e^{i \hat{p}_1 L / \hbar} | \psi_B \rangle$ survives despite the spatial non-overlap of $ | \psi_A \ra $ and $ | \psi_B \ra $. This is because the displacement operator $e^{i \hat{p}_1 L / \hbar}$ is non-local in the position basis. Therefore, $\langle e^{i \hat{p}_1 L / \hbar} \rangle$ exhibits a clear dependence on $\alpha$. This simple case establishes that for distinguishable particles, a \emph{single-particle modular operator} (acting solely on the Hilbert space of the particle that carries the phase) is a sufficient phase witness.

For indistinguishable particles, the initial state must be properly (anti)symmetrized for identical bosons and fermions, corresponding to Bose--Einstein (BE) and Fermi--Dirac (FD) statistics. With the defined single-particle states we have
\begin{equation} \label{eq: Psi0-indis}
| \Psi(0) \ra_{\pm} = N_{\pm} \left( \frac{ | \psi_A \ra_1 + e^{i\alpha} | \psi_B \ra_1 }{ \sqrt{2} } \otimes | \chi \ra_2 
\pm | \chi \ra_1 \otimes \frac{ | \psi_A \ra_2 + e^{i\alpha} | \psi_B \ra_2 }{ \sqrt{2} } \right) ,
\end{equation}
where
\begin{equation} \label{eq: nor-con}
N_{\pm} = \frac{1}{ \sqrt{ 2 \pm | \la \chi | \psi_A \ra + e^{i\al} \la \chi | \psi_B \ra |^2 ) } } ,
\end{equation}
is the normalization constant; and $+$ and $-$ stand for the symmetric (BE) and antisymmetric (FD) cases, respectively. From the corresponding density matrix,
\begin{eqnarray} \label{eq: rho0-indis}
\rho_{\pm}(0) &=& | \Psi(0) \ra_{\pm}~_{\pm}\la \Psi(0) |,
\end{eqnarray}
the reduced single-particle density matrix for either particle (tracing out the other) is:
\begin{eqnarray} \label{eq: rhosp0-indis}
\varrho_{\pm} &=& N_{\pm}^2 \bigg[ \frac{ | \psi_A \ra + e^{i\alpha} | \psi_B \ra }{ \sqrt{2} } \frac{ \la \psi_A | + e^{-i\alpha} \la \psi_B | }{ \sqrt{2} } + | \chi \ra \la \chi | 
\nonumber \\ 
&\pm& \left( \frac{ \la \psi_A | \chi \ra + e^{-i\alpha} \la \psi_B | \chi \ra }{ \sqrt{2} }~~
\frac{ | \psi_A \ra + e^{i\alpha} | \psi_B \ra }{ \sqrt{2} } \la \chi| + \text{h.c.}  \right) \bigg],
\end{eqnarray}
where ``h.c.'' stands for the Hermitian conjugate of the preceding term.
The expectation value of a single-particle translation operator for this reduced state becomes:
\begin{eqnarray}
\langle  e^{ i \hat{p} L / \hb } \rangle_{\pm} &=&  N_{\pm}^2 \bigg[ \frac{1}{2} e^{i\al} \la \psi_A | e^{ i \hat{p} L / \hb } | \psi_B \ra
\nonumber \\ 
&\pm& \left( \frac{ \la \psi_A | \chi \ra + e^{-i\alpha} \la \psi_B | \chi \ra }{ \sqrt{2} }~~
\frac{ \la \chi | e^{ i \hat{p} L / \hb } | \psi_A \ra + e^{i\alpha} \la \chi | e^{ i \hat{p} L / \hb } | \psi_B \ra }{ \sqrt{2} }  + \text{c.c.}  \right) \bigg],
\end{eqnarray}
where ``c.c.'' denotes the complex conjugate of the first term inside the parenthesis, and we have used the fact that states $ | \psi_A \ra $ and $ | \psi_B \ra $ are non-overlapping.

If the state \(|\chi\rangle\) has no spatial overlap with any of the states \(|\psi_A\rangle\) and \(|\psi_B\rangle\), then the results for both BE and FD statistics coincide and equal half of the value obtained for MB statistics. 
We now examine two specific cases where the state $ | \chi \ra $ is either $ | \psi_A \ra $ or $ | \psi_B \ra $. In the first case, $| \chi \ra = | \psi_A \ra $, we obtain
\begin{eqnarray}
\langle  e^{ i \hat{p} L / \hb } \rangle &=& \frac{1}{2} 
\begin{cases}
\frac{1}{\sqrt{3}} (2 e^{i\al} \la \psi_A | e^{ i \hat{p} L / \hb } | \psi_B \ra + e^{-i\al} \la \psi_A | e^{ i \hat{p} L / \hb } | \psi_B \ra^* ) & \text{bosons} 
\\
- e^{-i\al} \la \psi_A | e^{ i \hat{p} L / \hb } | \psi_B \ra^* & \text{fermions}. 
\end{cases}
\end{eqnarray}
In the second case, $ | \chi \ra = | \psi_B \ra $, we obtain
\begin{eqnarray}
\langle  e^{ i \hat{p} L / \hb } \rangle &=& \frac{1}{2} e^{i\al} \la \psi_A | e^{ i \hat{p} L / \hb } | \psi_B \ra
\begin{cases}
\frac{1}{\sqrt{3}} & \text{(bosons)}, \\
1 & \text{(fermions)}.
\end{cases}
\end{eqnarray}
This result shows that the expectation value of the modular variable coincides for MB and FD statistics, while it is reduced by a factor of $1/\sqrt{3}$ for BE statistics in this specific configuration.
Note that, in both cases, Eq.~\eqref{eq: nor-con} yields $N_+ = 1/\sqrt{3}$ and $N_- = 1$.

The above analysis confirms that the introduction of quantum statistics does not necessitate a transition from local to global modular observables. The phase sensitivity originates from a fundamental property: modular variables are intrinsically nonlocal. An operator such as $e^{ i \hat{p} L / \hbar }$ probes the wave function across spatially separated regions, enabling the detection of interference terms $\langle \psi_A | e^{ i \hat{p} L / \hbar } | \psi_B \rangle$ that remain inaccessible to observables local or diagonal in position. Consequently, whether the system involves a single particle or multiple distinguishable particles, the relative phase encoded in the single-particle wave functions can, in principle, be revealed through a local modular measurement on an appropriate subsystem. 

Motivated by this structural robustness, we now turn to the dynamical regime. In the remainder of this section, we consider a two-particle system obeying classical MB statistics and investigate the time evolution of modular variables under the double CL equation with a common environment, which generates environment-induced correlations between the particles. We conclude with a brief qualitative discussion of the limitations of standard coherence and entanglement quantifiers in detecting the relative phase.

\subsection{Two-particle Caldeira-Leggett dynamics}

Consider a two-particle system initially prepared in the separable state
\begin{eqnarray}
\Psi(x_1, x_2, 0)
&=& \frac{ \phi_A(x_1, 0) + e^{i \alpha} \phi_B(x_1, 0) }{\sqrt{2}}
\phi_A(x_2, 0),
\end{eqnarray}
and coupled to a common environment with temperature $T$ and relaxation coefficient $\ga$. The dynamics is governed by the two-particle CL master equation with a common bath. Expressed in relative and center-of-mass coordinates $ (r_1, R_1; r_2, R_2; t) $ where $ r_i = x_i - x'_i $ and $R_i = (x_i+x'_i)/2$, the master equation in the presence of a uniform gravitational field reads \cite{CaMoPo-PA-2010, Mo-EPJP-2025}
\begin{eqnarray} \label{eq: CL-common-1}
\frac{\pa \rho}{\pa t} &=&  \bigg[ \sum_{n=1}^2  \left( i \frac{\hb}{m} \frac{\pa^2}{\pa R_n r_n} - 2 \ga (r_1+r_2) \frac{\pa}{\pa r_n} \right) - \frac{D}{\hb^2} (r_1+r_2)^2 + \frac{mg}{i \hb} (r_1+r_2) \bigg]\rho(r_1, R_1; r_2, R_2; t),
\end{eqnarray}
where as before $D=2m\gamma k_B T$ is the diffusion coefficient.
Tracing over the second particle yields the reduced state of the first particle. 
As long as the wave packets stay non-overlapping, the expectation value of the modular variable takes the form
\begin{eqnarray} \label{eq: expval-mod-red}
\la \cos(\hat{p}_1L/\hb) \otimes \mathds{1}_2 \ra
&=& \frac{1}{2}
\exp \left[
- \frac{D L^2}{4\hb^2 \ga} e^{-4\ga t} \sinh(4\ga t)
- \frac{L^2}{4\si_0^2} e^{-4\ga t} \sinh^2(2\ga t)
- \frac{k^2 \si_0^2}{2}
\right] \nonumber \\
&\times&
\cos \left[
\alpha - L \frac{ 1-e^{-4\ga t} }{4}
\left( k + \frac{mg}{\hb \ga} \right)
\right].
\end{eqnarray}
In the Schr\"odinger limit this expression reduces to
$ e^{-k^2\si_0^2/2} \cos(\alpha - m g L t / \hb)/2 $.

A comparison of Eqs.~\eqref{eq: expval-mod-1p} and \eqref{eq: expval-mod-red} reveals the distinctive effect of a common environment. The modular variable of the first particle is inherently sensitive to the relative phase $\alpha$ due to its initial superposition. The environment, however, qualitatively alters this sensitivity through bath-induced correlations: it modifies the amplitude damping, oscillation frequency, and decoherence rate of the modular signal. By contrast, the expectation value
$ \la \mathds{1}_1 \otimes \cos(\hat{p}_2L/\hb) \ra $
is of order $ e^{-L^2 / 8 \si_0^2} $ and is therefore negligible.

It should be emphasized that the analytical expression in Eq.~\eqref{eq: expval-mod-red} is strictly valid only while the two wave packets remain spatially non-overlapping. Within the early-time regime $\gamma t\ll 1$, the expression clarifies the observed dynamics, as it factorizes into the form $\mathcal{A}(t)\cos[\Phi(t)]$. The amplitude behaves as
\begin{equation}
\mathcal{A}(t)\approx \frac{1}{2}e^{-k^2\sigma_0^2/2}\exp\!\left[
-\frac{D L^2}{\hbar^2}\,t
- L^2\gamma\!\left(\frac{\gamma}{\sigma_0^2}+\frac{4D}{\hbar^2}\right)t^2
\right],
\end{equation}
where the leading contribution in the exponent is linear in $t$ and proportional to the temperature, and therefore dominates the exponential damping. The constant prefactor $e^{-k^2\sigma_0^2/2}$, with $k$ being the central wave number of the right-moving packet, only fixes the overall scale of the signal. The phase is
\begin{equation}
\Phi(t)=\alpha+\Delta\phi(t),\qquad 
\Delta\phi(t)\approx -\left(Lk\,\gamma+\frac{mgL}{\hbar}\right)t
\qquad (\gamma t\ll 1),
\end{equation}
so that the initial oscillation frequency is
\begin{equation}
\omega_0 = L \left( k\,\gamma+\frac{mg}{\hbar} \right),
\end{equation}
containing a dissipative contribution and a gravitational contribution. Importantly, the gravitational field $g$ influences only the phase evolution of the modular expectation value determining both its oscillation frequency and its time-dependent variation, while the exponential decay of its amplitude is governed solely by thermal noise. Hence, the gravitational imprint can be extracted from the phase dynamics, independently of the thermal damping.

\subsection{Failure of Standard Coherence and Entanglement Quantifiers to Detect Relative Phase}
\label{subsec:nogo_coherence}

In this subsection, we explain why commonly used quantifiers of quantum coherence and entanglement are fundamentally insensitive to the relative phase between spatially non-overlapping wave packets, even in the presence of dissipation and environment-induced correlations. This insensitivity reflects a structural limitation rather than a model-dependent or parameter-dependent effect.


Quantum coherence measures are often defined through off-diagonal elements of the density matrix. In the present context, however, the key limitation is not merely diagonality but locality. Standard coherence quantifiers formulated in the position representation probe spatial correlations within local or quasi-local regions. For spatially non-overlapping wave packets, the phase-dependent cross terms connect components with disjoint spatial support and therefore do not contribute to such locally defined measures. Consequently, conventional coherence quantifiers, such as the $\ell_1$ norm of coherence evaluated in the position basis, $ C_{\ell_1}(\rho) = \int dr \int dR ~ | \rho(r, R, t)| $, and related notions of spatial coherence length, remain insensitive to the relative phase $\alpha$ within the non-overlapping regime considered here.
Entropy-based quantities exhibit the same limitation. The linear entropy, which quantifies the mixedness of a quantum state, depends only on traces of powers of the density matrix and is therefore invariant under changes in $\alpha$. This invariance persists under dissipative evolution governed by the CL master equation: although environmental noise suppresses the cross terms dynamically, it does not convert the relative phase into a locally accessible quantity.

The situation remains unchanged in bipartite continuous-variable systems. Standard entanglement measures for Gaussian states are fully determined by the covariance matrix, whose elements consist of expectation values of local position and momentum operators, or linear combinations thereof. Since these operators are local, their expectation values are insensitive to the relative phase encoded in cross terms connecting spatially disjoint components in the non-overlapping regime. Consequently, entanglement quantifiers derived from the covariance matrix, including those based on symplectic eigenvalues, do not capture the relative phase $\alpha$ in the regime of negligible spatial overlap, even when entanglement is dynamically generated through coupling to a common environment.

Importantly, this insensitivity is not accidental. It originates from a structural limitation of quantities derived from the density matrix through observables that are local in position space. 
Although the density matrix formally contains phase-dependent cross terms, this information is encoded in its off-diagonal structure and is not accessible through standard local observables in the non-overlapping regime, where the corresponding wave packets have disjoint spatial support. 
As a result, quantities constructed from locally accessible density-matrix elements remain insensitive to the relative phase. This limitation is therefore not a breakdown of the density matrix formalism itself, but rather reflects the restricted sensitivity of conventional observables. It is precisely this situation that motivates the introduction of modular variables. 
These type of operators explicitly connect spatially separated regions and directly probe the cross terms that encode relative phase information. They therefore represent a distinct class of observables capable of accessing nonlocal quantum information that is not captured by conventional coherence and entanglement quantifiers.

\section{Numerical Results} \label{sec: num-res}

For the numerical analysis we work in dimensionless units with \(m=\hbar=1\). The initial state consists of two Gaussian wave packets that are well localized and symmetrically placed around the origin at \(x=\pm L/2\), with \(L=50\). Both packets share the same initial width \(\sigma_0=1\). The right packet is given a kick momentum \(k=0.1\), while the left packet is initially at rest. The uniform gravitational field is chosen to be \(g=-3\).
Although different values of the relaxation rate and temperature are used in the numerical calculations, the maximum damping rate is $\gamma = 0.1$ and the minimum temperature is $T = 2$. Therefore, the condition $ k_B T \gg \hbar \gamma $, which is required for the validity of the CL equation, is well satisfied throughout.

For a Gaussian distribution with center \(x_c\) and width \(s\), the probability contained in the interval \([x_c-5s,\;x_c+5s]\) is \(0.99999\). Accordingly, we take this interval as the effective support of the wave packet, and treat \(x_c\pm 5s\) as the endpoints of its tails.

Although the initial wave packets are non-overlapping, their intrinsic spreading eventually leads to overlap. To determine the time at which this occurs, we impose an operational non-overlap condition based on the separation of their effective supports. The condition requires that the right tail of the left packet remains to the left of the left tail of the right packet. In the Schr\"odinger framework this translates into
\[
- \frac{L}{2} - \frac{1}{2} g t^2 + 5\sigma_t \;<\; \frac{L}{2} + \frac{\hbar k}{m}t - \frac{1}{2} g t^2 - 5\sigma_t .
\]
This inequality reduces to \(-L - \hbar k t/m + 10\sigma_t < 0\), which for our chosen parameters yields \(t<10.002\). Thus, working at times shorter than \(t=10.002\) ensures that the wave packets remain non-overlapping.

Within the CL framework, the non-overlap condition takes the form
\[
- \frac{L}{2} - \frac{g}{2\gamma}\bigl(t-\tau(t)\bigr) + 5w_t
\;<\;
\frac{L}{2} - \frac{\hbar k}{m}\tau(t) - \frac{g}{2\gamma}\bigl(t-\tau(t)\bigr) - 5w_t .
\]
For $\gamma = 0.001$ and $T = 2$, the inequality is satisfied for times $ t < 9.606 $. Keeping the same relaxation rate but increasing the temperature to $T = 15$, the non-overlap condition is reduced to $ t < 7.858 $. Finally, for $T = 15$ and $\gamma = 0.01$, the condition becomes $ t < 10.73 $.

In what follows we restrict the analysis to \(t\leq 2\), which safely lies within the non-overlapping regime.

\vspace{1cm}

\begin{figure} 
\centering
\includegraphics[width=15cm,angle=-0]{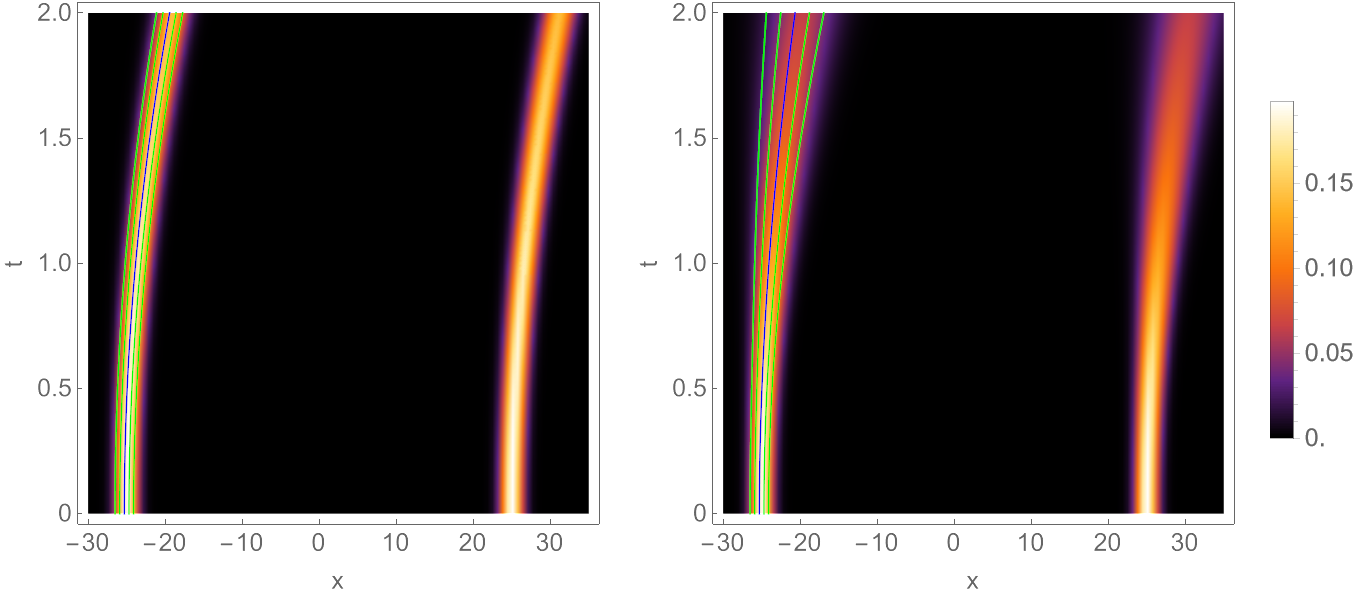}
\caption{
Density plots of the probability density for the Schr\"odinger dynamics (left) and the CL dynamics (right) at $\gamma = 0.1$ and $T=10$. A representative set of Bohmian trajectories is superimposed. These trajectories follow the motion of the left wave packet, while the right wave packet remains empty. The middle blue trajectory corresponds to the wave-packet center. The remaining parameters are fixed as $\hbar = m = k_B = \sigma_0 = 1$, $k=0.1$, $L=50$, and $g=-3$.
}
\label{fig: delplotTraj} 
\end{figure}

Figure~\ref{fig: delplotTraj} displays density plots of the probability density alongside representative Bohmian trajectories for both unitary Schr\"odinger dynamics and dissipative dynamics within the CL framework. In the left panel, the trajectories evolve according to Eq.~\eqref{eq: BM-traj-Sch}, where $\sigma_t$ (Eq.~\eqref{eq: sigmat}) governs the quantum-mechanical wave-packet spreading, $x_t$ is the classical center-of-mass motion, and $X_0$ is the initial Bohmian position. Consequently, trajectories with different initial positions $X_0$ diverge as $\sigma_t$ grows in time. The right panel shows the corresponding CL evolution for $\gamma=0.1$ and $T=10$. Here, the trajectories follow Eq.~\eqref{eq: BM-traj-CL}, with the effective width $w_t$ (given by Eq.~ \eqref{eq: width-CL}) incorporating both quantum spreading and thermal diffusion ($D\propto T$). This leads to a markedly faster and broader dispersion of trajectories compared to the unitary case. Furthermore, the center-of-mass motion $x_t$ is modified by friction: the contribution from the initial momentum is scaled by the rescaled time $\tau(t)$, which saturates to $1/(2\gamma)$ for $t \gg \gamma^{-1}$. After a transient period, the entire bundle of trajectories therefore drifts with a \emph{terminal velocity} $v_{\mathrm{term}} = -g/(2\gamma)$, in contrast to the constantly accelerated motion in the Schr\"odinger dynamics. In both panels, only trajectories belonging to the left (occupied) wave packet are displayed; the right packet remains empty. The middle (blue) trajectory, which starts at the packet center ($X_0=x_0$), follows $x_t$, while the surrounding trajectories diverge in accordance with the wave-packet broadening. This visualization highlights how environmental coupling amplifies the separation between neighboring trajectories and thereby accelerates the loss of spatial coherence.

\begin{figure} 
\centering
\includegraphics[width=12cm,angle=-0]{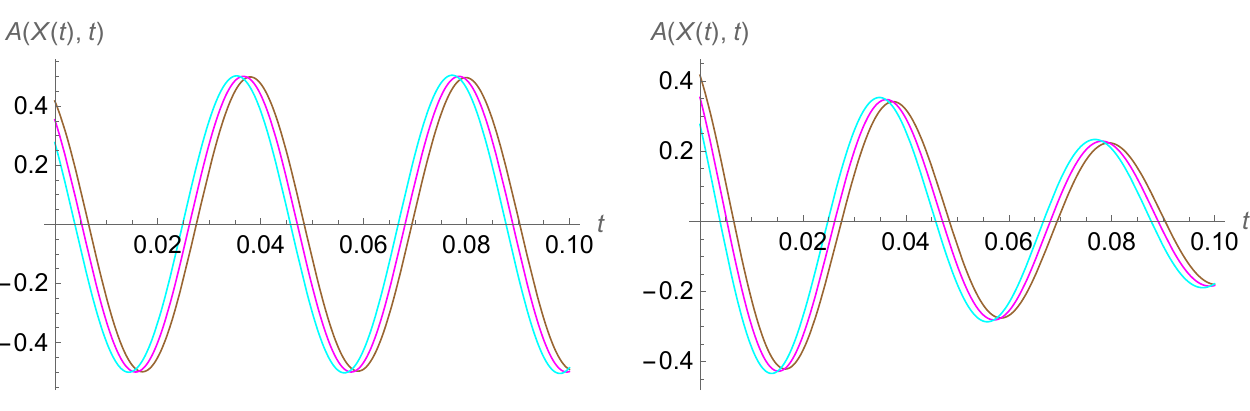}
\caption{
Local expectation value of the modular variable along Bohmian trajectories for a relative phase $\alpha = \pi/4$ and three different Bohmian initial positions: $X_0 = -L/2 - 2\sigma_0$ (brown), $X_0 = -L/2$ (magenta), and $X_0 = -L/2 + 2\sigma_0$ (cyan). The left panel shows the Schr\"odinger dynamics [Eq.~\eqref{eq: expval-traj}], and the right panel shows the CL dynamics. The remaining parameters are $\hbar = m = k_B = \sigma_0 = 1$, $k = 0.1$, $L = 50$, $g = -3$, $\gamma = 0.001$, and $T = 2$.
}
\label{fig: pAloctrajCL_X0} 
\end{figure}

Figure \ref{fig: pAloctrajCL_X0} displays the local expectation value $A(X(t),t)$ of the Hermitian modular operator evaluated along individual Bohmian trajectories for a fixed relative phase $\alpha=\pi/4$ and three different initial Bohmian positions $X_0$, as determined by Eq.~\eqref{eq: expval-traj}. The left panel corresponds to purely unitary Schr\"odinger evolution, for which the analytic expression \eqref{eq: expval-traj} applies exactly. In this case, $A(X(t),t)$ exhibits a pronounced oscillatory behavior in time, governed by the cosine factor whose argument contains contributions from the relative phase $\alpha$, the gravitationally induced dynamical phase proportional to $mgLt/\hb$, and trajectory-dependent phase shifts proportional to $k X_0 \sigma_0/\sigma_t$. These contributions lead to clear phase shifts between the oscillations associated with different initial positions $X_0$, while preserving a common oscillation frequency set primarily by the gravitational and momentum-dependent terms. The exponential prefactor in Eq.~\eqref{eq: expval-traj}, which depends on the wave-packet spreading through $\sigma_t$ defined in Eq.~\eqref{eq: sigmat}, remains close to unity for the chosen parameters and time interval, resulting in oscillations of nearly constant amplitude. The right panel shows the corresponding results obtained from a solution of the CL equation. These results reveal a clear qualitative modification of the unitary behavior: while the oscillatory structure and the relative phase shifts between different Bohmian trajectories persist, the amplitude of the modular signal decreases with time due to environmental dissipation and thermal noise. Importantly, the damping affects all trajectories in a similar manner, indicating that the environment primarily suppresses the modular signal through amplitude decay rather than by washing out the phase dependence. This comparison highlights the robustness of the phase information encoded in local modular expectation values, even when evaluated along individual Bohmian trajectories in an open-system setting.

\begin{figure} 
\centering
\includegraphics[width=12cm,angle=-0]{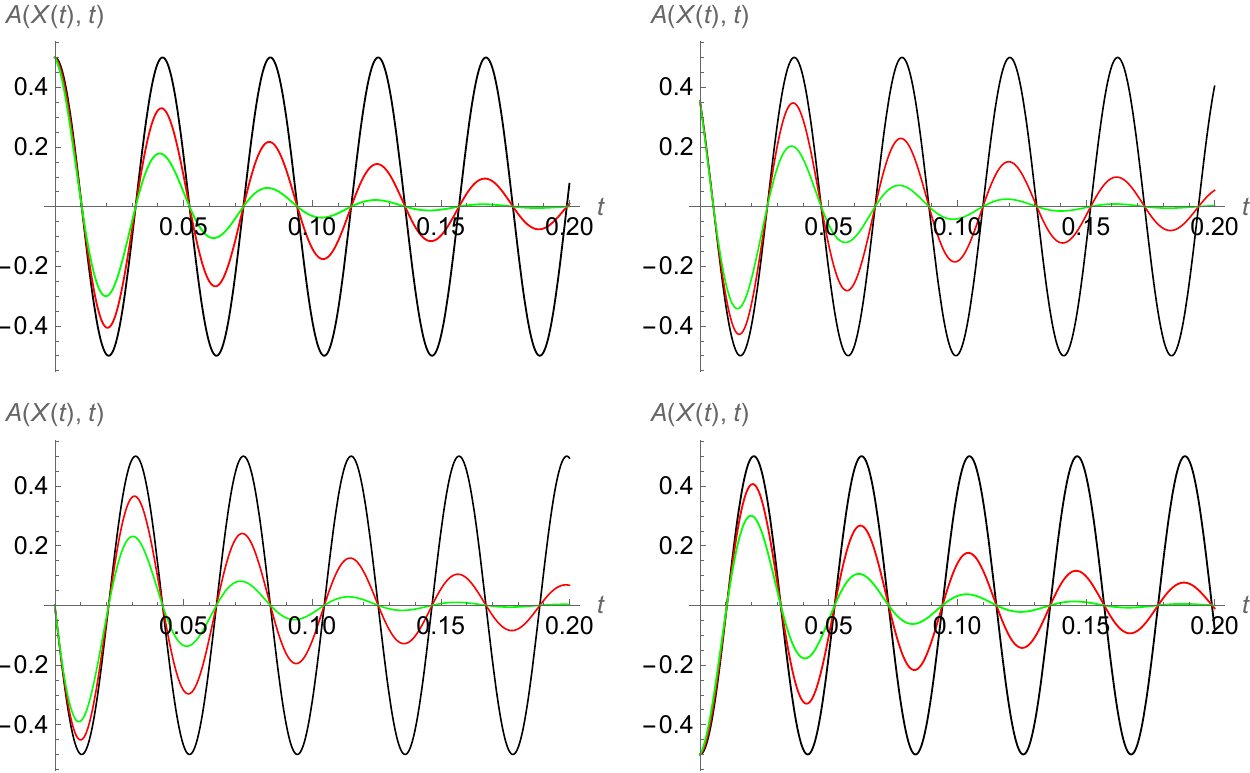}
\caption{
Local expectation value of modular variable along the Bohmian trajectory with initial position $X_0 = -L/2$, shown for different values of the relative phase $\alpha$: $\alpha = 0$ (top left), $\alpha = \pi/4$ (top right), $\alpha = \pi/2$ (bottom left), and $\alpha = \pi$ (bottom right). In each panel, the black curve corresponds to the Schr\"odinger dynamics [Eq.~\eqref{eq: expval-traj}], while the red and green curves correspond to the CL framework with temperatures $T = 2$ and $T = 5$, respectively, and a relaxation rate $\gamma = 0.001$. The remaining parameters are $\hbar = m = k_B = \sigma_0 = 1$, $k = 0.1$, $L = 50$, and $g = -3$.
}
\label{fig: pAloctrajCL} 
\end{figure}

Figure \ref{fig: pAloctrajCL} illustrates the local expectation value of the modular operator evaluated along a Bohmian trajectory with fixed initial position $X_0=-L/2$, for four different values of the relative phase $\alpha$. The panels correspond to $\alpha=0$ (top left), $\alpha=\pi/4$ (top right), $\alpha=\pi/2$ (bottom left), and $\alpha=\pi$ (bottom right). In each panel, the black curve represents the unitary Schr\"odinger evolution, while the red and green curves correspond to solutions of the CL equation for temperatures $T=2$ and $T=5$, respectively. The comparison of the four panels, each associated with a different value of $\alpha$ within the same dynamical regime, clearly reveals the dependence of the modular expectation value on the relative phase. Under unitary dynamics, the modular expectation value exhibits persistent oscillations whose relative displacement depends explicitly on $\alpha$. As $\alpha$ is varied, the oscillatory pattern undergoes a clear shift while its amplitude and frequency remain essentially unchanged, indicating that the relative phase primarily enters through the argument of the cosine rather than through the envelope. In contrast, the dissipative dynamics induced by the environment leads to a gradual suppression of the oscillation amplitude, with stronger damping observed at higher temperature. Importantly, although environmental noise reduces the visibility of the modular signal, the relative phase dependence remains clearly distinguishable at intermediate times, as evidenced by the systematic displacement of the oscillatory patterns between panels. This behavior highlights that, while the environment decoheres the modular expectation value through amplitude damping, it does not eliminate the underlying phase sensitivity encoded in the modular operator, thereby reinforcing the role of modular variables as carriers of nonlocal phase information in open quantum systems.

\begin{figure} 
\centering
\includegraphics[width=12cm,angle=-0]{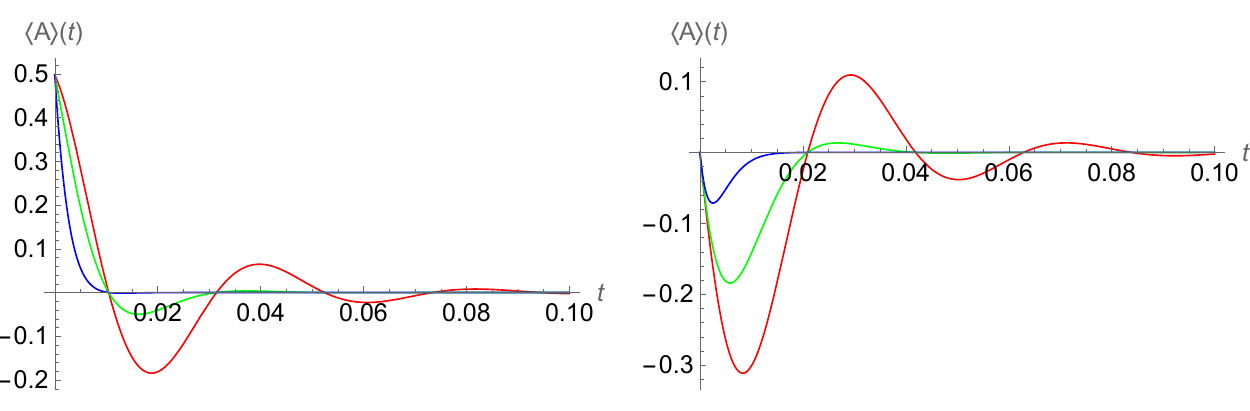}
\caption{
Expectation value of the modular variable of the first particle [Eq.~\eqref{eq: expval-mod-red}] for relative phases $\alpha=0$ (left panel) and $\alpha=\pi/2$ (right panel). Colored curves correspond to the CL framework with relaxation
rate $\gamma=0.005$ and temperatures $T=2$ (red), $T=5$ (green), and $T=15$ (blue).
Other parameters are $\hbar=m=k_B=\sigma_0=1$, $k=0.1$, $g=-3$ and $L=50$.
}
\label{fig: pAexpCLreduced} 
\end{figure}

Figure~\ref{fig: pAexpCLreduced} displays the time evolution of the expectation value of the modular variable of the first particle, $\langle \cos(\hat{p}_1 L/\hbar) \otimes \mathds{1}_2 \rangle$, as obtained from Eq.~\eqref{eq: expval-mod-red}. The evolution is shown only for the finite time interval in which the two wave-packet components, $\rho_1$ and $\rho_2$, that constitute the reduced density matrix of the first particle (see Appendix) remain spatially non-overlapping, thereby ensuring the validity of the analytical expression. The figure shows this
quantity as a function of time for two representative values of the relative phase,
$\alpha=0$ (left panel) and $\alpha=\pi/2$ (right panel), within the CL framework
at a fixed relaxation rate $\gamma$, and for three different temperatures
$T=2$ (red curves), $T=5$ (green curves), and $T=15$ (blue curves).
The dynamics is characterized by oscillations whose phase evolution is governed by the argument
of the cosine term, which depends explicitly on the relaxation rate $\gamma$ and therefore
reflects the influence of the environment on the modular dynamics. At the same time, the cosine
argument is independent of the temperature $T$, implying that thermal noise does not alter the
phase-sensitive structure of the oscillations. In contrast, temperature enters solely through
the exponential prefactor via the diffusion coefficient $D\propto T$, which controls the damping
of the oscillation amplitude. As a consequence, increasing $T$ leads to a faster suppression of
the oscillation visibility, causing the modular signal to be washed out at earlier times, while
leaving the underlying phase evolution intact. This separation between dissipative phase
modification (through $\gamma$) and thermal decoherence (through $D$) shows that, within the
regime of non-overlapping wave packets, thermal noise acts primarily by suppressing the
visibility of modular oscillations, while the phase-sensitive dynamics encoded in modular
variables remains directly influenced by environmental dissipation.

\section{Summary and Conclusions} \label{sec: sum-con}

In this work, we have investigated the role of modular variables as probes of nonlocal phase information in superpositions of spatially non-overlapping wave packets. The central result of our analysis is that, although the density matrix of such superpositions formally contains phase-dependent cross terms, these contributions remain operationally inaccessible through observables that are local or quasi-local in position space. As a consequence, conventional quantum-mechanical quantities including standard coherence measures, entropy-based indicators, and entanglement quantifiers, are structurally insensitive to the relative phase as long as the wave packets remain non-overlapping. This limitation is independent of whether the system evolves unitarily or undergoes dissipative open-system dynamics.

Modular variables overcome this limitation by explicitly connecting spatially separated regions of the wave function. Through analytical and numerical investigations, we demonstrated that modular expectation values retain direct sensitivity to the relative phase even in regimes where conventional observables fail. 
In particular, we showed that gravitational evolution induces a characteristic time dependence in modular signals that encodes phase information, while probability density and current, although dynamically affected by gravity, remain phase-insensitive in the non-overlapping regime.
We note that the conclusions drawn here are not restricted to the specific case of a linear potential, but are expected to extend to more general potentials due to the nonlocal structure $V(x+\ell)-V(x)$ governing the dynamics of modular variables. Moreover, in the presence of significant overlap between wave packets, standard local observables already become sensitive to the relative phase, and the role of modular variables is correspondingly less essential.

Using local expectation values within the Bohmian framework, we developed a trajectory-resolved description of modular observables, providing a clear dynamical interpretation of their phase sensitivity. This analysis clarifies the conceptual status of empty waves: although dynamically inert when spatial overlap is negligible, empty waves carry nonlocal quantum information that becomes operationally accessible through modular operators. Importantly, this informational role emerges without modifying the underlying Bohmian dynamics and remains fully consistent with standard quantum-mechanical predictions.

We further extended the analysis to two-particle systems, demonstrating that the phase-detection capability of modular variables persists in the presence of quantum statistics and environment-induced correlations. In particular, for particles coupled to a common environment, we showed that environment-induced entanglement modifies the modular signal associated with one particle, providing a direct signature of environmental influence. At the same time, the transfer of phase sensitivity to the second particle produces only a negligible effect in its corresponding modular signal. These findings establish a clear operational distinction between local quantum correlations and nonlocal phase information encoded in modular observables.

Our open-system analysis reveals that thermal noise solely damps the amplitude of modular observables, while the phase information encoded in their oscillatory argument is unaffected by temperature; only the dissipation rate $\gamma$ influences the phase evolution. Environmental decoherence therefore limits the detectability of modular observables through amplitude damping rather than phase randomization. The phase in the cosine argument depends linearly on the relative phase $\alpha$ as $\Phi(t) = \alpha + \Delta\phi(t)$, where $\Delta\phi(t)$ is influenced by dissipation but not by thermal noise. Thus, the $\alpha$-dependence remains exact and is obscured only by the decay of the amplitude.
It should be noted that, since the CL equation is an effective description valid under the assumptions of weak coupling, Markovian dynamics, and sufficiently high temperature, these results should be interpreted within this regime of validity.

Finally, we emphasize that Bohmian mechanics serves here as an interpretational framework providing additional dynamical insight rather than new empirical predictions. Since Bohmian trajectories are not directly observable for individual particles, the local expectation values evaluated along these trajectories should be understood as tools for clarifying the physical structure underlying standard quantum theory. Within this perspective, our results illuminate how nonlocal quantum information can be encoded in empty-wave components while remaining hidden from local dynamical observables.

Taken together, our findings establish modular variables as a robust framework for probing nonlocal phase information in both isolated and open quantum systems, and clarify the interplay between nonlocality, decoherence, and dynamical evolution in quantum mechanics.

\vspace{2cm}

\noindent
{\bf Acknowledgement}:
Support from the university of Qom is acknowledged. 
We express our gratitude to the referees for their thorough review and constructive suggestions.

\vspace{0.5cm}
\noindent
{\bf Data availability}: This manuscript has no associated data.



\appendix
\section{Solution of the CL Equation for Superposed Gaussian Packets in a Gravitational Field} \label{sec: appendix}

In this appendix, we present the solution to the CL equation \eqref{eq: CL1p-rR} for the initial state given by \eqref{eq: rho0}. Since the evolution equation is linear in the density matrix, each term in the initial density matrix can be evolved independently and then superposed to obtain the final solution. Thus, we have
\begin{equation}
\rho(r, R, t) = \frac{1}{2} \big( \rho_1(r, R, t) + \rho_2(r, R, t) + e^{i\al} \rho_3(r, R, t) + e^{-i\al} \rho_4(r, R, t) \big),
\end{equation}
where
\begin{align*}
\rho_1(r, R, 0) &= \psi_A(R/2+r, 0)\,\psi^*_A(R/2-r, 0), \\
\rho_2(r, R, 0) &= \psi_B(R/2+r, 0)\,\psi^*_B(R/2-r, 0), \\
\rho_3(r, R, 0) &= \psi_B(R/2+r, 0)\,\psi^*_A(R/2-r, 0), \\
\rho_4(r, R, 0) &= \psi_A(R/2+r, 0)\,\psi^*_B(R/2-r, 0).
\end{align*}
Using the method of characteristics, we find that each term shares the same functional form, given by
\begin{eqnarray} \label{eq: func-form}
\rho_{j}(r, R, t) &=& \frac{1}{\sqrt{2\pi} w_t} \exp \left[ a_j(r, t) - \frac{( R + i ~ b_j(r, t) )^2}{ 2 w_t^2 } \right]
\end{eqnarray}
with the following coefficients:
\begin{eqnarray}
a_1(r, t) &=& -\frac{i m g  \left( 1 - e^{-2 \ga  t} \right)}{2 \ga  \hb} r + \frac{e^{-4 \ga  t} \left(-2 D \si_0^2 \left(e^{4 \ga  t}-1\right)-\ga  \hb^2\right)}{8 \ga  \hb^2 \si_0^2} r^2 , \\
b_1(r, t) &=& \frac{i \left(-2 \ga  g t-g e^{-2 \ga  t}+g-2 \ga ^2 L\right)}{4 \ga ^2}
-\frac{e^{-4 \ga  t} \left(e^{2 \ga  t}-1\right) \left(2 D \si_0^2 \left(e^{2 \ga  t}-1\right)+\ga  \hb^2\right)}{8 \ga ^2 \hb m \si_0^2}r ,
\\
a_2(r, t) &=& a_1(r, t) + i k r e^{-2 \ga  t} ,
\\
b_2(r, t) &=& b_1(r, t) + i \left( L + \frac{h k \left(1-e^{-2 \gamma  t}\right)}{2\gamma  m} \right) ,
\\
a_3(r, t) &=& a_1(r, t) - \frac{4 k^2 \si_0^4 + 4 i k L \si_0^2 + L^2}{8 \si_0^2} + \frac{e^{-2 \ga  t} \left(L+2 i k \si_0^2\right)}{4 \si_0^2} r ,
\\
b_3(r, t) &=& b_1(r, t) + \frac{e^{-2 \ga  t} \left(L+2 i k \si_0^2\right) \left(-\hb+e^{2 \ga  t} \left(\hb+4 i \ga  m \si_0^2\right)\right)}{8 \ga  m \si_0^2} ,
\\
a_4(r, t) &=& a_3(r, t) + L \left( i k - \frac{e^{-2 \gamma  t}}{2 \si_0^2} r \right) ,
\\
b_4(r, t) &=& b_3(r, t) + 2 k \si_0^2 + \frac{\hb L \left(e^{-2 \gamma  t}-1\right)}{4 \gamma  m \si_0^2} .
\end{eqnarray}

\end{document}